CSCE Annual Conference

*Growing with youth – Croître avec les jeunes*

Laval (Greater Montreal)

June 12 - 15, 2019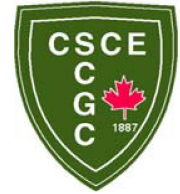

# ZERO LATENCY FOR EMERGENCIES: A MACHINE LEARNING BASED APPROACH TO QUANTIFY IMPACT OF CONSTRUCTION PROJECTS ON EMERGENCY RESPONSE IN URBAN SETTINGS

**Zhengbo Zou[1] and Semiha Ergan[2]**

[1]PhD Candidate, Department of Civil and Urban Engineering, New York University, 15 MetroTech Center, Brooklyn, NY, 11201; email: zz1658@nyu.edu
[2]Assistant Professor, Department of Civil and Urban Engineering, New York University, 15 MetroTech Center, Brooklyn, NY, 11201; email: semiha@nyu.edu**Abstract:** Continuous construction and rehabilitation in urban settings have unavoidable impacts on arrival times of first responders to emergency locations, which play an essential role in the quality of life for residents. Current research efforts on emergency response assessments focus on case studies, where specific periods (e.g., 9/11 and storm Sandy) of emergency response times are analyzed. Simulation based studies that aim to evaluate response times in relation to various constraints/fleet sizes also exist. Although such research work can help in studying the response times in disruption periods or with resource constraints, they do not analyze how specific changes (e.g., new and ongoing construction projects) in urban settings impact emergency response times of first responders. This paper aims to fill the gap and proposes a novel approach to predict the expected emergency response time for a given location using the fabric of zones regarding construction activities. This approach relies on historical records of emergency response and construction permits issued by city agencies, where the log of construction activities around neighborhoods are maintained. The approach first defines the signature of a zone (defined by zip codes) for construction activities based on the distribution of historical construction work types permitted in that zone over time. Then, zones that share similar signatures are clustered to find if there exists a relationship between construction signatures and emergency response times. Next, supervised learning algorithms are deployed to predict the average emergency response times for each cluster. The approach was tested using New York City's construction permit and emergency response records. This study can be replicated in other cities using the proposed approach to predict and compare travel times of first responders in zones in a city using the fabric of zones regarding construction activities. This study serves as the first step towards quantitatively understanding construction projects' impact on a quality of life (QoL) indicator (specifically emergency response times) in urban settings, which can be extended for other QoL indicators.## 1    INTRODUCTION

Ongoing construction and rehabilitation are common scenes in densely packed cities. From 2013 to 2017, New York City Department of Buildings (DOB) issued an average of 500 permits per day for building construction projects (DOB 2018). Construction activities have indisputable impacts on residents' quality of life, especially for emergency response (i.e., fire, police, and medical services). The Fire Department of New York (FDNY) reported more than 2.7 million incidents regarding fire and medical emergencies in the same four-year period (FDNY 2017). A direct relation of construction and emergency response is the travel time from a dispatch location to an incident location. Navigating in dense cities with congested and fully/partially closed streets due to construction work increases the travel time for first responders.

Construction fencing and enlarged construction perimeter also contribute to these delays in rescue times (Peraza 2007). In 2016, FDNY reported an average response time of more than 11 minutes in the densely packed borough of Manhattan for medical emergencies; 9 minutes for the Brooklyn and Bronx; 8 minutes for Queens; and 7 minutes for Staten Island (FDNY 2017). These numbers also fluctuate over zones in a given borough. Therefore, these numbers are hard to improve unless a full understanding of the distribution of types of construction activities is gained for each location. However, despite the anticipated impact (i.e., delay) of construction projects on emergency response, the on-and-off nature of construction work and how it changes the habitat of regions have not been analyzed at a large scale yet. City agencies need to be informed about contributors of the fluctuating response times across zones (when resources and vicinities are controlled) to suggest improvements for these response times.

Currently, the understanding of the construction projects' impact on nearby buildings' emergency response time is limited (Zou and Ergan 2019). Typically, first responders are dispatched to an incident location based on their availabilities and the vicinity of the incident location to the dispatching unit. The central dispatch officers also maintain a database containing ongoing construction works that are close to dispatch locations (Challands 2010, Zhong et. al., 2010). Additionally, any construction activities that require road closures or traffic obstructions need to apply for a permit from the Department of Transportation (NYC DOT 2018), and notify the fire/police stations nearby (DOB 2008). Despite these efforts of trying to keep the first responders in the loop for ongoing construction activities in a vicinity, the city agencies (FDNY, DOT and DOB) still lack the ability to fully understand the distribution of construction works happening on a larger scale and their influence in variations of response times in similar capacity zones. Furthermore, the city agencies are also in need of a predictive approach that can indicate how the ongoing construction activities influence the emergency response times at given locations, to take proactive actions to mitigate the potential delaying effect.

This paper aims to fill the gap and proposes a novel approach to predict the expected emergency response times given the historical distribution of construction activities in a zone. The idea is to inform the city agencies of the distribution of ongoing construction activities on an urban scale, and then provide a quantitative method with the goal of anticipating the potential delaying effect of the construction activities in various locations. This approach relies on emergency response records and historical records of construction permits issued by city agencies, where log of construction activities around neighborhoods are maintained. The proposed machine learning based approach includes two steps: (1) calculating signatures of each zone of a city regarding construction work and find clusters of construction signatures using unsupervised clustering algorithms, and (2) predicting emergency response times using the construction signatures and clustering results from step one, and the historical emergency response data. The approach was tested using New York City's construction permit data records of around 900 thousand and emergency response logs of around 2.7 million from 2013 to 2017. The analysis was kept in 2013-2017, where both datasets were available.

We consider the contribution of this paper essential because: (1) this paper is the first known attempt to quantify building construction projects' impact on the emergency response time on an urban scale, regardless of type of construction activities (e.g., foundation, new buildings, demolition); and (2) this paper proposed a novel machine learning based approach to analyze large open city data, which can be used for further analysis of construction projects' impact on the residents' quality of life from other perspectives (e.g., traffic and environment).

## 2  BACKGROUND

Existing studies of emergency response analyze response times during disruptive events (Veijalainen and Hara 2011). Research studies on emergency evacuation plan and triage allocation during disruptive times, such as when twin towers collapsed are among such studies (Tierney 2003, De Goede 2008). Super storm Sandy also caused another wave of research focusing on the flooded roads' influence on emergency response times (Comes and Van de Walle 2014). Another example is the wild fire in California, which sparked research studies looking into the emergency response plans for combatting strong wind coupled with wild fires (Cohn et. al., 2006, Keeley et. al., 2004). Although such work can help in studying the response plans for withstanding the conditions imposed by specific extreme events, they do not shed light on emergency response during routine times, and particularly how ongoing construction activities in urban settings impact emergency response times of first responders.

Emergency response is also an active research focus in the operational research domain, where much of previous studies can be described as works assuming the availability of well-structured emergency response services (Simpson and Hancock 2009). To elaborate, first responders' resources and processes are defined with the assumption that required service is fulfilled by a single station. These assumptions also include the type of emergencies. Given these assumptions, the likelihood of a unit's availability for a given incident can be analyzed (Yuan and Wang 2009). To use fire emergency as an example, the response time is assumed to be directly related to the number of fire stations, the distance between the fire station and the incident location, and ignores any other conditions that can impact the service (Kolesar and Blum 1973). There are also simulation-based studies that aim to evaluate response times in relation to various constraints/fleet sizes (Caunhye et. al., 2012). For example, in the area of disaster relief operational studies, researchers control variables such as the resource availability, connectivity of the road network, and the distance to the incident locations to optimize the emergency response times (Erkut and Polat 1992). However, there is still a potential to perfect the response system for the purpose of minimizing the travel delay where past incidents' dispatch data are analyzed (Altay et. al., 2006).

In conclusion, although the literature is quite heavy in the operational research domain for the optimization of emergency response times based on a set of constraints, such as resource availability, distance of incidents, and road network connectivity, there is no study in the literature that examined the impact of distribution of construction activities at an urban scale on emergency response times. As a result, city agencies are not equipped with analysis tools that will reveal the pattern of construction activities in city zones and enable them to minimize fluctuations in response times in different zones. One of the challenges for analyzing the impact of construction projects in urban level services is the unavailability of structured historical data. In recent years, the broader availability of public city data regarding construction projects (e.g., location and duration, scope, and construction work type) and emergency response (e.g., incident location and time, response unit, and response time) has provided opportunities to quantify the influence of construction projects on emergency response times at a large scale (Zou and Sha 2018).

## 3   DATA OVERVIEW

The approach introduced in this paper uses the open city data published by the New York City (NYC) Department of Buildings (DOB) and the Fire Department (FDNY) from 2013 to 2017. DOB regulates "more than one million buildings and active construction sites in New York City by enforcing construction laws, building codes, and zoning resolutions" (DOB 2018). FDNY responds to fire and medical emergencies in the five boroughs of New York City. These datasets were used to evaluate the clustering and prediction performance of the approach. It is essential to introduce the datasets earlier in this paper to better comprehend the examples used in explaining the approach, as the examples are specific to these datasets.

### 3.1   Construction Permit Data from NYC Department of Buildings

NYC DOB issues permits for construction activities regulated by the NYC construction laws and building codes. The permits are accessible online through DOB's open data portal (DOB 2018). The available permit data ranges from 1983 to 2017. The data is in a tabular format, with about 3.5 million rows and 60 columns. The essential columns included in the data analysis include three categories of information: (1) location information (i.e., borough, zip code, street, and house number); (2) work information (i.e., work number, work type, and work subtype); and (3) time information (i.e., permit start date and permit expiration date). The location and time information are self-explanatory. The work information, especially work type and work subtype, is one of the main parameters describing the nature of construction projects that this study builds on. The full list of the work types and the work subtypes (if any) in the dataset is shown in Figure 1.

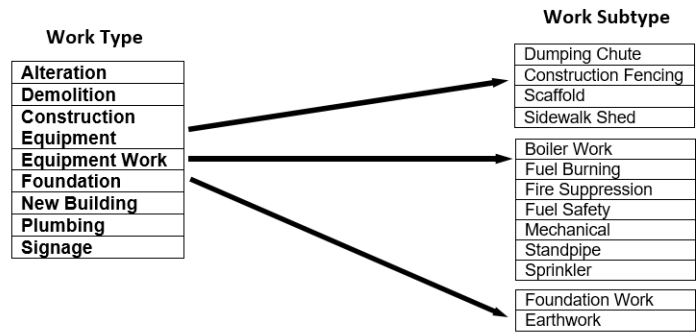

Figure 1. Construction Work Types and Work Subtypes included in DOB Permit Data

From work types and work subtypes, we can categorize the construction activities based on the possible effect they have on the emergency response times (i.e., interior works have minimal influence on the emergency response times because they are self-contained). Among the eight work types, construction equipment (including dumping chute, construction fencing, scaffold and sidewalk shed), foundation (including foundation work and earthwork), and new buildings have the potential to request for partial/full closures and can potentially influence the emergency response times as compared to the other work types.

### 3.2 Emergency Response Data from the Fire Department of New York City

FNDY responds to the fire and medical emergencies in NYC. Every incident reported is recorded with the incident location, time, and response time. The data is in a tabular format with 2.7 million rows and 29 columns documenting fire incidents from 2013 to 2017. The columns used in the data analysis include two categories of information, namely, (1) location information (i.e., incident borough and zip code) and (2) time information (i.e., incident date, time, and incident response time). Incident response time is defined as the time elapsed (in seconds) for the first responding unit to arrive at the incident location.

## 4 OVERVIEW OF THE APPROACH AND EVALUATION OF RESULTS USING NYC OPEN DATA

The proposed approach includes two main steps. The first step is to find the construction signatures to understand the characteristics of construction activities for each zone (i.e., zip code) in a city. Here, a *construction signature* is defined as the distribution of the construction work types within a given zip code. This is achieved by calculating the percentage of each work type (e.g., foundation, plumbing) happened in a specific zip code using the construction permit dataset. The signatures are then clustered based on zip codes using unsupervised learning algorithms to find clusters of zip codes that display a similar construction signature. The underlying assumption is that zip codes with similar construction signatures will show similar emergency response observations (i.e., the response times are distinctly different among clusters). The input for this step is the construction projects database (e.g., DOB permit data), which includes information such as *120 5$^{th}$ Ave, Manhattan, 10014, alteration, start date 01/01/2016, end date: 01/01/2017*. The output for this step is the construction signature and the cluster number for each zip code (e.g., *10001, 3.8% alterations, 14.3% construction equipment, …, 13.5% new building, cluster number 2*). We initially started with the k-means algorithm in this step for clustering, however this initial work will be extended to evaluate the performance of other clustering algorithms to identify the high performing algorithm for this problem.

The second step of the approach is to create prediction models of the independent variables (i.e., construction signature for each zip code with the associated cluster number, and the corresponding emergency response time for each zip code) and the dependent variable (i.e., the average emergency response time) using supervised learning algorithms. The goal of this step is to provide quantitative insights of construction projects' impact on emergency response times by allowing the estimation of the response time given the knowledge of construction permit signatures of each zip code. The input for this step is the construction signatures obtained from the first step and the emergency response database (e.g., FDNY emergency response dataset), which includes information such as *Manhattan, 10014, incident datetime 02/09/2017 3pm, fire incident, response time 312 seconds*). The output is the predicted average emergency response time given a zip code and its cluster number (e.g., *for zip code 10002 which belongs to cluster 2, the average response time is 368 seconds*). We used three algorithms (i.e., ordinary least squares, decision tree, and random forest) and will be extending this work to evaluate the performance of various other algorithms.

## 4.1 Signatures of DOB Construction Permit Data

We defined the notion of '*construction signature*', using Equations 1 and 2, to understand the unique characteristics of the construction activities distributed across different zip codes of NYC. Let the construction work type be $t$, and the zip code be $z$, then the total number of permits issued for work type $t$ within zip code $z$ is defined as $s(t, z)$. The signature of zip code $z$ can be represented using a vector $S(z)$:

$$S(z) = \frac{s(t,z)}{s(z)}, t = 1, 2, \ldots, T. \quad (1)$$

$$where, s(z) = \sum_t s(t, z), t = 1, 2, \ldots, T \quad (2)$$

In the above equation, $s(z)$ is the total number of permits in zip code $z$, and $T$ is the total number of permitted work types. The vector $S$ highlights the primary permitted work types in the specific zip code, allowing a straightforward comparison across zip codes and cities. Table 1 shows two examples of construction signatures for two zip codes in NYC.

Table 1: Example Construction Signatures: *S(10001) and S(10002)* for Zip Codes 10001 and 10002

| Zip Code | New Building | Foundation Work | Construction Equipment | Demolition | Alteration | Equipment Work | Plumbing | Signage |
|---|---|---|---|---|---|---|---|---|
| 10001 | 1.6% | 1.7% | 14.3% | 0.5% | 3.8% | 62.2% | 13.5% | 3.4% |
| 10002 | 2.3% | 1.7% | 16.1% | 0.9% | 5.3% | 50.5% | 21.6% | 1.6% |

A construction signature is unique to the zip code it is defined for, and shows the distribution of the historical construction activities in a given zip code. The hypothesis is that the differences in the construction signatures (i.e., the distribution of the construction work types in zip codes) can be used to infer the emergency response time for zip codes. To test this hypothesis, we first applied unsupervised learning algorithms to cluster the calculated construction signatures around zip codes (results shown in Figure 2). Then, we compared the emergency response times of first responders across clusters (results shown in Table 2).

We chose k-means clustering algorithm in this initial work, since it has shown accurate clustering results (when compared to other clustering algorithms) for large open city data (Wang et. al., 2017). The k-means algorithm was run 100 times (a typical number used to eliminate the bias introduced by randomly initializing the starting weights in the k-means algorithm) to find the optimal number of clusters. The metric used to compare the goodness-of-fit of the data points to the clusters was the Silhouette Score, which can be calculated using Equation 3. Intuitively, the Silhouette Score measures how similar a data point is to its own cluster compared to other clusters (i.e., how well a cluster represents that data point given the other points in that cluster). Therefore, a larger Silhouette Score means a better clustering result. Let $c(i)$ be the average distance between data point $i$ and all other data points within the same cluster (intra cluster distances). Let $o(i)$ be the smallest of the average distances of $i$ to all points in any other cluster other than the cluster $i$ is assigned to (inter cluster distances). Silhouette Score $s(i)$ is defined as:

$$s(i) = \frac{o(i) - c(i)}{\max(c(i), o(i))} \qquad (3)$$

We ran the algorithm with the number of clusters ranging from 2 to 100. The best Silhouette Score was achieved when the number of clusters was set to 5. The clusters are drawn in Figure 2 overlaying the NYC zip code map. The construction signatures for the five clusters (i.e., the average percentages of the construction work types for each cluster) are also shown in Figure 2.

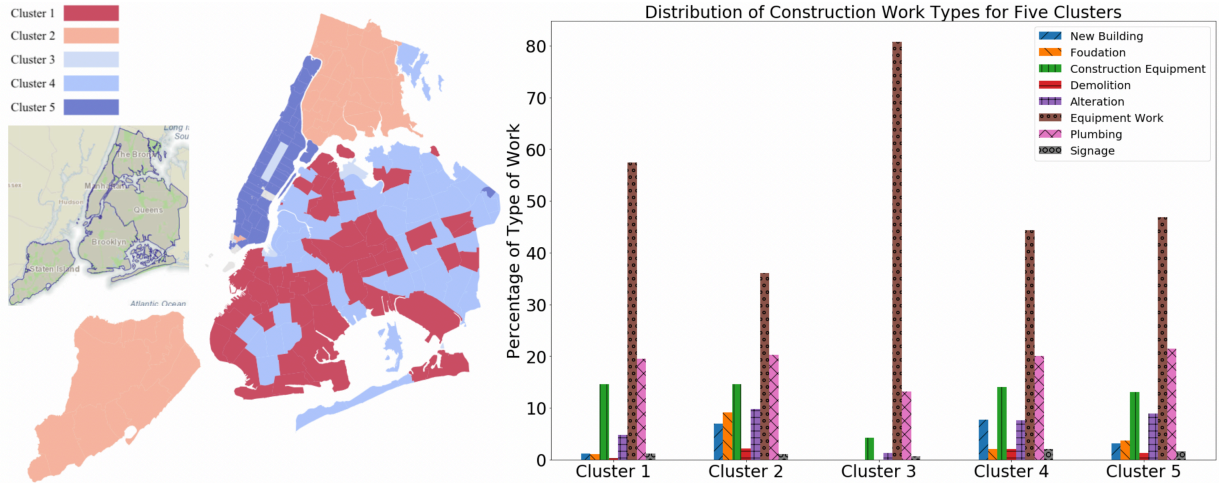

Figure 2. Clusters of Construction Signatures and Distribution of Work Types within Clusters

Figure 2 shows the similarities and distinctions among clusters in terms of the distribution of the work types. First, the common theme throughout NYC, regardless of clusters, is the dominance of equipment work (see Figure 1 for details of what this type contains). On the other hand, the clusters do also differ in other types of the construction activities. For example, Cluster 3, which is where natural reserves and parks are covered (see the NYC map above), has the most distinct distribution of construction work with respect to the other clusters with the highest equipment use, no new construction/demolition, and minimum construction equipment outside. It is also the cluster of parks and recreational areas where the minimum incidents (3% of the total incidents) were reported with respect to the other clusters. Hence, Cluster 3 was removed from further analysis. The rest of the clusters are similar in terms of construction equipment usage, but differ in amount of new building, foundation, demolition, foundation work. Cluster 2 differs in construction fabric with respect to others for the amount of foundation work. For Cluster 2 and 4, which are mostly residential areas and rural suburbs, new buildings and demolition are relatively higher than the other clusters. Cluster 1 has the smallest alteration, foundation and new buildings work in its distribution (given that Cluster 3 is set aside). Cluster 5 resembles to Cluster 4 in its construction signature except the new construction work, which is expected to change the response times in these regions.

To understand if the construction signature clusters can help explain differences in travel times of first responders to incident locations, the authors looked at the average emergency response time in the four remaining clusters, as shown in Table 2.

Table 2. Average Emergency Response Time for Clusters

| Cluster | 1 | 2 | 4 | 5 |
|---|---|---|---|---|
| Average Response Time (seconds) | 288 | 354 | 343 | 304 |

The results show that for Clusters 2 and 4, where the percentage of new building and demolition works are dominant, the average response times for emergencies are higher than other clusters. This concurs with the assumption that exterior works (e.g., new building and demolition works) require a larger construction premise, where construction fences and sidewalk sheds need to be erected, hence deterring the traffic within those locations, and slowing down the emergency response time. Also, foundation work is higher in the signature of Cluster 2 with respect to Cluster 4, which might be contributing to higher response times in Cluster 2. Cluster 1 has the shortest response time, which is expected given it has the smallest alteration, foundation and new buildings works in its distribution. Cluster 5 resembles to Cluster 4 in its construction signature, except for the new construction work (being lower), which is expected to change the response times in these regions. Response time of Cluster 5 being smaller than Cluster 4 supports our hypothesis.

The observations above reveal support for the hypothesis of construction signature and its impact on emergency response times. While the clustering was performed using construction permit data alone, the average response time for emergencies happen to be distinctive among clusters. However, the clustering results only provide evidence that there is a relationship between construction fabric of a region and the response times in that region. The next step in the data analysis approach is to determine whether construction signatures can be used to predict the average emergency response time, given a location.

## 4.2 Predicting Average Emergency Response Times in Clustered Regions using Construction Signatures

The goal of this prediction is to inform the city agencies (e.g., Fire Department of New York City) of the potential influence from the construction projects happening in the vicinity by looking into the predicted average emergency response times. The authors used supervised learning methods, specifically, regression models for this step. Ordinary Least Squares was used in this step for benchmarking. Decision Tree and Random Forest were used due to their high accuracies achieved when modeling open city data (Wang et. al., 2017). Ordinary Least Squares (OLS) is one of the simplest ways of implementing linear regression (Dempster et. al., 1977), which is commonly used as a benchmarking model for other machine learning methods to compare to. Because the assumption of linearity is almost certainly too aggressive for urban data, OLS should produce the least accurate result, and serve as a baseline for other methods.

Decision Tree (DT) algorithm is based on the Tree Structure. The branches carry weights that represent the probabilities of the previous nodes propagating to that branch. The regression value is a weighted aggregated value of all leaf nodes. Random Forest (RF) was also used in this approach because of its capability of being used without the need for extensively large datasets or long training times. Earlier studies that compared most widely used supervised learning techniques on empirical datasets showed that Random Forest algorithm is the best performer (Caruana and Niculescu-Mizil 2006). The reason for the high accuracy of the RF algorithm is that it creates a collection of base prediction models, and use the aggregation of base models as the final prediction result, which eliminates the individual modeling biases.

The input of the regression models was the construction signatures of clusters from step one. The prediction target for the regression models is the average emergency response time aggregated based on zip codes. Cross validation was used to tune the parameters of the models, and the best performing model was selected based on the R-Squared value, which is commonly regarded as the first choice of benchmarking parameter to use (Salakhutdinov et. al., 2007). Models with high R-Squared value have higher probability of correctly predicting unseen data. The results of the R-Squared value for the three selected models are shown in Table 3.  Cluster 3 was again excluded for its small sample size.

Table 3: Regression Model Performance Measured by R Squared for Average Emergency Response Time

| Cluster | Ordinary Least Squares | Decision Tree | Random Forest |
|---------|------------------------|---------------|---------------|
| 1 | 0.59 | 0.72 | 0.76 |
| 2 | 0.50 | 0.67 | 0.77 |
| 4 | 0.52 | 0.79 | 0.81 |
| 5 | 0.54 | 0.65 | 0.72 |

One observation from Table 3 is that Random Forest showed the highest R-Squared value in all clusters for predicting the average response time, comparing to the lowest results achieved by the benchmarking algorithm of Ordinary Least Square for all clusters. OLS assumes linearity for the input, which is an oversimplification for urban datasets, and caused underfitting. Random Forest achieved better performance than Decision Tree because it created multiple base models during the training process, and can better handle the unseen values than Decision Tree.

The results of the regression models are considered essential since they showed the possibility of using construction signatures as an input to predict the emergency response data. For example, the city agencies (e.g., Fire Department or Department of Buildings) can use the regression models built using the historical data to predict the possible emergency response time for a zip code given its current construction signature (i.e., the distribution of construction work types). Finally, the regression methods proposed in this study can be easily scaled by geographic aggregation for other cities, and for other types of open city data.

## 5  CONCLUSION AND FUTURE WORK

This paper fills the gap of existing emergency response research by proposing a quantitative approach for the city agencies to predict the expected emergency response time given the historical distribution of construction activities in a location. This approach relies on historical records of construction permits issued by city agencies, where log of construction activities around neighborhoods are maintained as well as emergency response recordings. The authors showed the possibility of using construction signatures (i.e., distribution of the construction work types at a given location) to create clusters of zones with distinct construction signatures, where the emergency response time also differs. The authors also showed the possibility of building prediction models using construction signatures to predict the emergency response time. This paper serves as the first step towards understanding how construction fabric in a given location can influence the nearby emergency response services as compared to other locations. City agencies and the first responders can use the proposed approach to understand the impact of construction projects for emergency response on an urban scale regardless of the building types or the construction work types. The next step of this work would be to test other algorithms to evaluate the algorithms that better fit to the problem domain, and the addition of other open city data sources (e.g., DOT data), with the goal of understanding the construction impact on emergency response and other aspects of urban quality of life.